\documentclass[preprint]{aastex}

\shortauthors{Palla et al.}
\shorttitle{$^{12}$C/$^{13}C$ in NGC~3242}

\newcommand{\heh}{$^{3}$He/H~}
\newcommand{\crat}{$^{12}$C/$^{13}$C~}
\newcommand{\cdod}{$^{12}$C~}
\newcommand{\ctre}{$^{13}$C~}
\newcommand{\etre}{$^3$He~}
\newcommand{\ciii}{\ion{C}{3}]~}
\newcommand{\msun}{$M_\odot$}
\def\arcsecpoint{\ifmmode ''\!. \else $''\!.$\fi}

\begin{document}

\title{The \crat ratio in the Planetary Nebula NGC~3242 from 
{\it Hubble Space Telescope} STIS 
observations\altaffilmark{1}}
\altaffiltext{1}{Based on observations with the NASA/ESA {\it Hubble Space
Telescope}, obtained at the Space Telescope Science Institute, which is
operated by the Association of Universities for Research in Astronomy
(AURA), Inc., under NASA contract NAS~5--26555.}

\author{Francesco Palla, Daniele Galli, Alessandro Marconi}
\affil{INAF -- Osservatorio Astrofisico di Arcetri \\
Largo Enrico Fermi 5 \\
I-50125 Firenze, Italy}
\email{palla,galli,marconi@arcetri.astro.it}

\author{Letizia Stanghellini\altaffilmark{2}}
\affil{Space Telescope Science Institute\\
3700 San Martin Drive\\
Baltimore MD 21218, USA}
\email{lstanghe@stsci.edu}
\altaffiltext{2}{Affiliated with the Astrophysics Division, Space Science
Department of ESA; on leave from INAF--Osservatorio Astronomico di Bologna}

\and

\author{Monica Tosi}
\affil{INAF -- Osservatorio Astronomico di Bologna \\
Via Ranzani, 1\\
I-40127 Bologna, Italy}
\email{tosi@bo.astro.it}

\begin{abstract}
We present high resolution HST ultraviolet 
spectroscopy of NGC~3242,
the only planetary nebula with a measured abundance of $^3$He. 
The Space Telescope Imaging Spectrograph (STIS) has been used to observe
the C{\sc iii} multiplet near $\lambda 1908$~\AA. 
The presence of \cdod and \ctre lines at these wavelengths allows a direct 
estimate of the carbon isotopic ratio in the ionized gas of the nebula.
We have detected the \cdod doublet and obtained an upper limit on the
\ctre $\lambda$1909.6~\AA~line, resulting in a carbon isotopic ratio
$^{12}$C/$^{13}$C$>$38, in agreement with standard stellar models.
The lack of the \ctre line
and the presence of the \etre line in the spectrum indicate that the
progenitor star did not undergo a phase of deep mixing during the last
stages of its evolution. The significance of this result for studies of
stellar nucleosynthesis and Galactic chemical evolution is discussed.

\end{abstract}


\section{Introduction}

In principle, the abundance of $^3$He in our Galaxy can be used to
test the predictions of Big-Bang nucleosynthesis and to provide constraints 
on the baryon density of the Universe~\citep[e.g.][]{ban02}.
However, the usefulness of $^3$He as a {\em
cosmic baryometer} is limited by a poor understanding of its evolution
in our Galaxy, one of the major unsolved mysteries in the field of
cosmic abundances.  The problem was first recognized by ~\citet{roo76}
who noted the strong discrepancy between the theoretical yields of
low-mass stars, $X(^3{\rm He})\sim 10^{-3}$ for a 1~$M_\odot$
star~\citep{ibe65}, and the measured abundances in H{\sc ii} regions,
$X(^3{\rm He}) \sim 10^{-5}$~\citep{bal99a}.  Despite the increase
in the number and quality of observational constraints and in the
accuracy of stellar evolutionary calculations over the past 25 years,
the $^3$He problem is still with us today \citep[e.g.][]{tos01}.

An interesting solution to the puzzle of $^3$He based on stellar
physics has been suggested by various groups
\citep{hog95,cha95}.  Accordingly, a non-standard mixing
mechanism acting during the RGB and/or AGB phases of stars of mass up
to $\sim 2 M_\odot$ can effectively suppress the production of $^3$He.
In this way, low-mass stars are not $^3$He producers and may even
become net $^3$He destroyers, lessening the tension between the results
of chemical evolution models and observations.  If this mechanism is 
indeed at work in
low-mass stars, an unavoidable consequence is that the ratio of
$^{12}$C/$^{13}$C in the ejecta of planetary nebulae (PNe) 
should be much {\em lower} than in the
standard case. For a  $1 M_\odot$ star, the predicted ratio is $\sim 5$
against the standard value of 25--30 \citep{boo99}.  Thus, it is very
important to have an accurate measure of the carbon isotopic ratio in those
PNe where a high $^3$He abundance has been determined.  Should these
nebulae  show a $^{12}$C/$^{13}$C ratio close to 25--30, then no
modifications to the standard scenario would be required.  Otherwise,
one has to invoke another selective process (circulation, diffusion etc.)
that operates on $^{13}$C but not on $^3$He.

NGC~3242 is the only PN with a firm detection of the $^3$He$^+$
hyperfine line at 8.67 GHz. The initial discovery indicated a value of
\heh$\simeq 10^{-3}$ by number \citep{roo92}. Subsequent observations
confirmed the detection and
refined the estimate of the abundance to lower values,
\heh$=(2-5)\times 10^{-4}$ \citep{bal99b}.  In both cases, the
resulting values of \etre agree with the predictions of standard
stellar nucleosynthesis models \citep{roo76}.

In an attempt to derive the \crat ratio in PNe, we observed at
millimeter wavelengths a sample of 28 objects, determing the CO
isotopic ratio in 14 of them and obtaining robust upper limits in 6
other nebulae \citep{pal00}. Unfortunately, no CO emission was detected
in NGC~3242, so that we could not draw any conclusion about the
occurrence of mixing in this PN. Following \citet{cle97}, we decided to
adopt the complementary approach of observing the \ciii multiplet near
1908~\AA, exploiting the unique spectroscopic capabilities of HST/STIS at
ultraviolet wavelengths.  \citet{cle97} measured for the first time the
carbon isotopic ratio in the ionized gas of NGC~3918 and SMC~N2, using
the GHRS aboard the HST. 
The big advantage of this method over observations in the
millimeter band is that the simultaneous presence of the \cdod and \ctre
lines in this spectral region allows a {\it direct} estimate of the carbon
isotopic ratio without having to worry about problems such as isotopic
fractionation and optical depth effects that make the derivation of
abundances from CO lines more uncertain \citep{sah94,sch00}.

\section{Observations and Results}

NGC~3242 is a multiple-shell, attached halo PN, located at a distance
of about 1~kpc \citep{sta95}. The inner and outer shells have diameters of
15$''$ and 29$''$, respectively. The inner shell (or rim) is characterized by
the presence of two prominent spots in ~\ion{O}{1}, \ion{O}{3}, and
\ion{N}{2}.  The post-AGB mass of the central star is estimated to be 
$M_{\rm CS}=(0.56\pm 0.01)$~\msun, corresponding to a main-sequence mass of 
$M_{\rm MS}=(1.2\pm 0.2)$~\msun ~\citep{gal97}. The metallicity of NGC~3242
is $Z\simeq Z_\odot/2$~\citep{bar85}.

We present high quality STIS observations of \cdod and \ctre in the
ionized gas of NGC~3242 by using the \ciii multiplet near 1908~\AA.
The multiplet has two lines, $^3$P$^0_1$--$^1$S$_0$ at 1908.7~\AA~ and
$^3$P$^0_2$--$^1$S$_0$ at 1906.7~\AA~ which are intercombination and
magnetic quadrupole transitions, and a third line,
$_{1/2}^3$P$^0_0$--$_{1/2}^1$S$_0$ at 1909.6~\AA, which is generally
completely forbidden except in the case of the \ctre atom as a result
of the non zero nuclear spin. The first two lines have different
transition probabilities and their ratio has been used to determine the
electron density in PNe and other low-density objects.  The third line
is weaker than the others (typically, less than one percent), and
originates only in $^{13}$C.

We observed NGC~3242 with STIS on the {\it HST} over 5 orbits in March
2001 for a total of 13000 seconds of integration time.  All of the
spectra were obtained through a 52$''\times 0$\arcsecpoint1 slit to
maximize throughput without a significant loss in spectral resolution.
The slit was positioned to cut through the two conspicuous spots of the
rim, avoiding the central star. We used the MAMA detectors and G230M
gratings whose range well encompasses the spectral region of interest,
and whose pixel scale (0.09~\AA~per pixel) is adequate for our
observations.  The prime tilt was centered at 1933~\AA, achieving a
spectral resolution of 0.36~\AA~ ($\sim 25$~km~s$^{-1}$) which is
enough to resolve the \ciii line at 1909.6~\AA~ from the stronger
\ciii line at 1908.7~\AA.

The spectra were recalibrated using the STIS pipeline
and the best reference files available in the archive as of November 2001.
The individual spectra from each orbit were
independently calibrated in wavelength and flux, and resampled to a
linear wavelength scale (retaining the same average dispersion).
Careful examination of the spectra from each orbit reveals no
discernible change in flux over time or as a function of position on
the detectors. We therefore averaged the spectra (weighted by exposure
time) to obtain the final versions of the G230M spectra.
The spectrum was then obtained by coadding $\sim 250$ rows selected
from the highest surface brightness regions coinciding with
the two bright spots of the rim.

\begin{figure}
\plotone{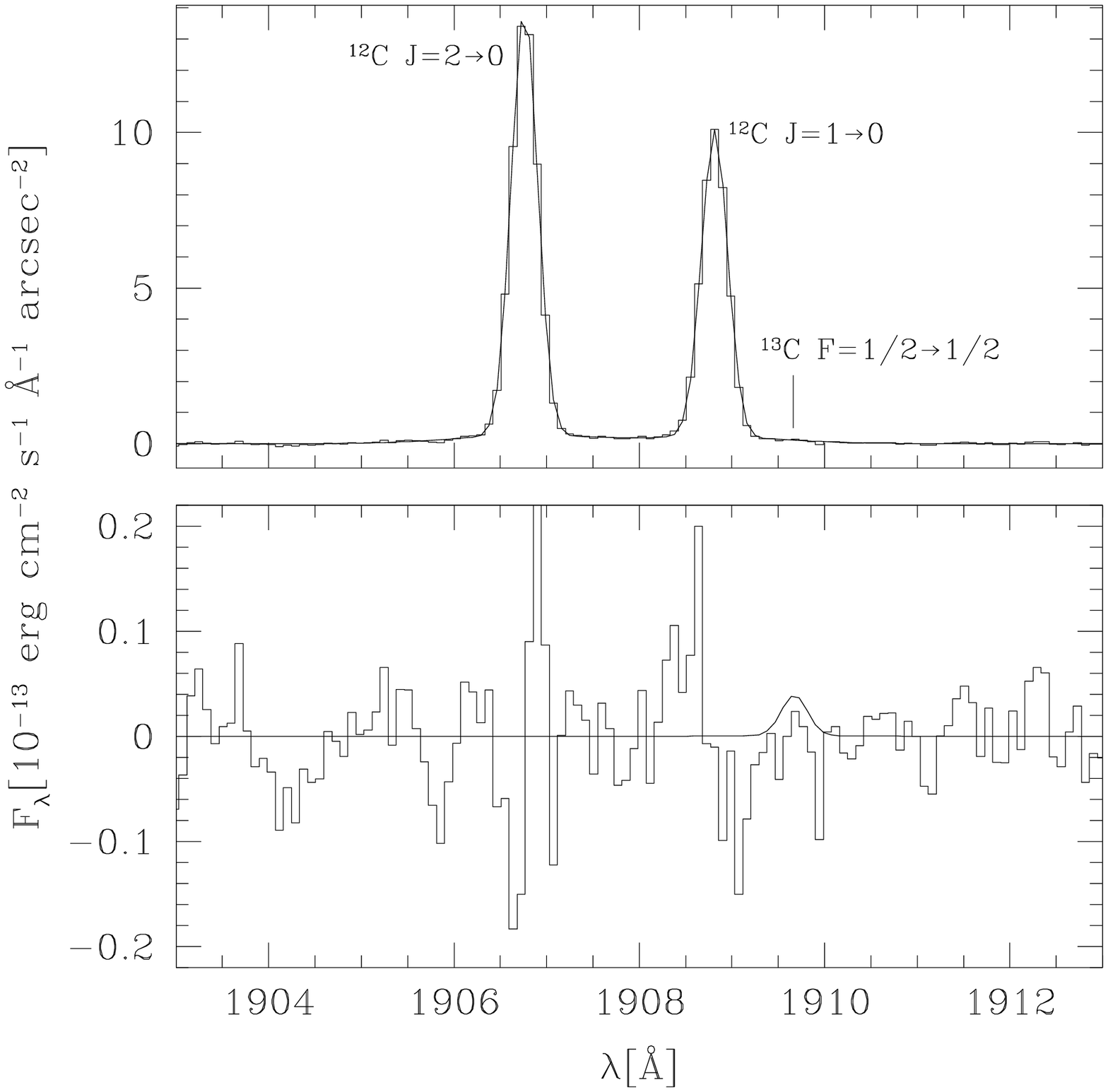}
\caption{HST-STIS spectra of NGC~3242. Upper panel: Observed spectrum
(histogram) of  the \ciii multiplet with the fit superimposed ({\em solid
line}). Lower panel: fit residuals; the {\em dotted line} represents a \ctre
line at 1909.6~\AA~ with a flux ratio of 0.003 relative to the 1906.7~\AA~
line. The \ctre line is not detected, thus setting the value of the firm 
upper limit \crat$>38$.}
\label{fig:fit}
\end{figure}

The resultant spectrum and its residuals are shown in Figure~\ref{fig:fit}.
The \cdod lines at 1906.7~\AA~ and 1908.7~\AA~ have fluxes of $(5.36\pm
0.04)\times 10^{-13}$ and $(3.91\pm 0.04)\times 10^{-13}$~erg cm$^{-2}$ s$^{-1}$
arcsec$^{-2}$, respectively, yielding a ratio of 1.37$\pm 0.02$.  For the
\ctre line at 1909.6~\AA, we have obtained a $1\sigma$ upper limit of $\sim
1.5\times 10^{-15}$~erg cm$^{-2}$ s$^{-1}$ arcsec$^{-2}$.  Thus, \ciii
1909.6/1906.7$<2.8\times 10^{-3}$.

The \cdod lines cannot be described with gaussians profiles because they
present faint blue and red wings. As shown in Table~1, we fit the
\cdod lines with two gaussian components with the same central wavelength
but different fluxes and line widths.  The bright narrow component has a FWHM
of 52 km s$^{-1}$, significantly larger than the instrumental resolution of
the G230M grating ($\sim 30$~km s$^{-1}$); the broadening is probably due to
the presence of two unresolved components coming from the receding and 
approaching edges of the nebula.  
The faint wings can be described with a gaussian profile with 
290~km s$^{-1}$~FWHM, indicating the presence of fast moving gas.

Following \citet{cle97}, the 1906.7/1908.7 intensity ratio yields the
electron density, while the 1909.6/1908.7 line ratio is used to determine the
\crat abundance ratio (see their Fig.~2).  For a value of
$1906.7/1908.7=1.4$, we obtain an electron density $n_e=8.9\times
10^3$~cm$^{-3}$. This value is larger than that reported by \citet{sta89}
from a variety of forbidden lines of lower ionization
($n_e=2.7\times 10^3$~cm$^{-3}$), but still below the threshold for 
collisional quenching of the
\ctre line ($n_e\simeq 2\times 10^4$~cm$^{-3}$). Then, the expected
1909.6/1908.7 ratio computed assuming equal abundances of \cdod and \ctre is
0.15. Using the observed limit on $1909.6/1906.7<2.8\times 10^{-3}$, we infer
a lower limit to the carbon isotopic ratio of \crat$>0.15/(1.4\times
2.8\times 10^{-3})\simeq 38$.

In Figure~\ref{fig:abund}, we compare the observed \heh \citep{bal99b} and
\crat ratio (this work) of NGC~3242 with the predictions of standard and 
non standard models for the advanced phases of the evolution of stars in the
mass range 1 to 3 $M_\odot$.  Note that the largest discrepancy between the
models occurs below $\sim 2 M_\odot$, i.e for stars that do
experience the helium flash. It is evident that standard models (left panels)
reproduce very well the results in both cases, whereas the inclusion of
extra-mixing processes fails by factors between 3 and 5. 
As noted above,
the metallicity of NGC~3242 is lower than solar by about a factor of 2.
For standard models, the \etre yields are rather insensitive to $Z$,
as shown by the calculations of \citet{dea96} (see their Fig.~1).
Similarly, a 50\% decrease of $Z$ produces only a $\sim$20\% increase
of the \crat ratio ~\citep[e.g.][]{van97}. 

Considering the sensitivity of the non-standard models to metallicity,  the
effect is shown in Fig.~\ref{fig:abund} where we plot two curves for $Z=0.02$
and $Z=0.007$. At lower metallicities, the depth of the convective zone is
larger, implying higher temperatures and more efficient nuclear burning at 
the base of the convection region.
Therefore, the predicted \etre abundance {\it and} \crat ratio are lower 
than in the solar case, and the discrepancy with the observed values 
becomes even larger.

\begin{deluxetable}{llll}
\tablewidth{0pt}
\tablecaption{\sc Line parameters of the \ciii multiplet}
\label{tab:line}
\tablehead{
\colhead{Line} &
        \colhead{$\lambda$\tablenotemark{a}} &
        \colhead{Flux\tablenotemark{b}} &
        \colhead{FWHM\tablenotemark{c}} \\ }
\startdata
\cdod $J=2\rightarrow 0$                 &  1906.760$\pm$0.001 & 4.85$\pm$0.03
 & 52.3$\pm$0.2 \\
                                         &  1906.760           & 0.51$\pm$0.03
& 290$\pm$25 \\
\cdod $J=1\rightarrow 0$                 &  1908.81 & 3.50$\pm$0.02 & 52.3 \\
                                         &  1908.81 & 0.41$\pm$0.03 & 290 \\
\ctre $F={1\over 2}\rightarrow {1\over 2}$ & 1909.67  & $<0.015$ &  \\
\enddata
\tablenotetext{a}{~Observed Wavelength (\AA).}
\tablenotetext{b}{~Surface brightness ($10^{-13}$~erg~cm$^{-2}$~s$^{-1}$ 
arcsec$^{-2}$).}
\tablenotetext{c}{~Full Width at Half Maximum (km~s$^{-1}$).}
\end{deluxetable}

\section{Discussion}

\begin{figure}
\plotone{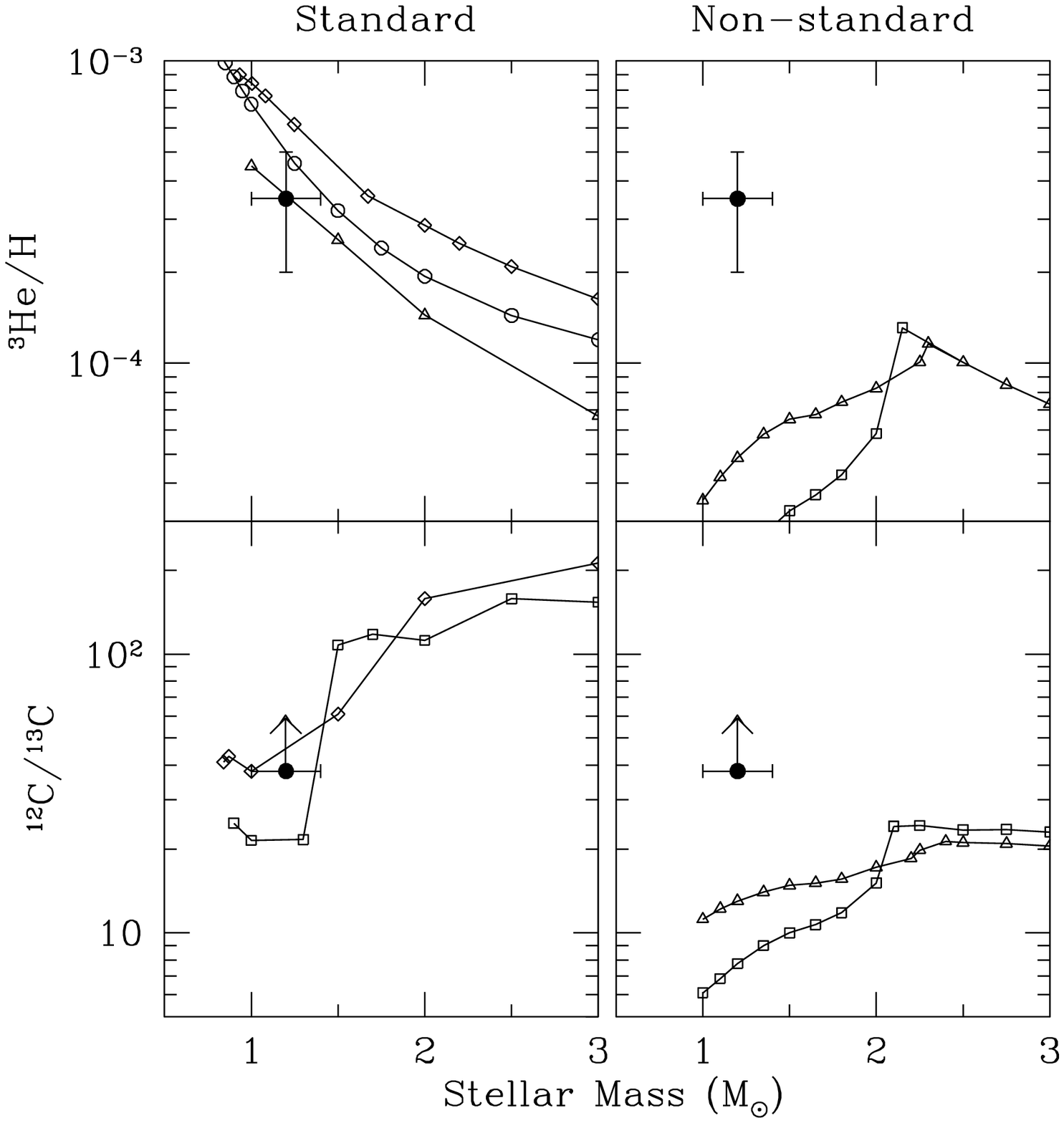}
\caption{Elemental abundances vs. progenitor mass for standard ({\it
left panels}) and non-standard ({\it right panels}) stellar evolution
models.  The observed $^3$He/H abundance of NGC~3242 \citep{bal99b} and
the lower limit on \crat found in this work are indicated by the {\em
filled dots}.  The mass estimate of NGC~3242 and associated
uncertainty are from \citet{gal97}.  {\it Standard models}: for \etre
the symbols show the models computed by \citet{mar01} ({\em diamonds}), 
\citet{wei96} ({\em triangles}), and~\citet{dea96} ({\em circles}); 
for the \crat ratio the symbols show the
models of \citet{van97} ({\em squares}) and ~\citet{mar01}
({\em diamonds}) for the ejecta of stars at the tip of the AGB phase.
All models are for solar metallicity.  {\it Non-standard models}: the
symbols show the results of \citet{sac99} for ~\etre ~and
~\citet{boo99} for ~\crat: $Z=0.02$ ({\it triangles}) and $Z=0.007$
({\it squares}) at the end of the RGB phase.
}
\label{fig:abund}
\end{figure}

The HST observations reported here provide an independent support to the 
\etre observations at 9.8 GHz and allow to draw firm conclusions about
the nucleosynthesis history of NGC~3242. 
The lack of the \ctre line
and the presence of the \etre line in the spectrum indicate that the
progenitor star did not undergo a phase of deep mixing during the late
stages of its evolution.  At first glance, this result may suggest that
there is no need to revise standard nucleosynthesis models, since
Fig.~\ref{fig:abund} shows that there is excellent agreement between
their predictions and the values derived from the observations of both
~\etre and ~\crat. However, the question is whether NGC~3242 is a special 
case among PNe.

The existence of isotopic anomalies in low-mass stars (below 
$\sim 3 M_\odot$) is well documented for red giant stars 
\citep[e.g.][]{char98,gra00}.  From the
analysis of a large sample of field and cluster RGB stars, \citet{nas98}
conclude that more than 90\% of them present \crat ratios inconsistent with
the results of standard models. The case for AGB stars is more
controversial.  Evidence for mixing comes from data on oxygen and carbon
isotopic ratios \citep{was95} and on $^7$Li abundances \citep{cha00}, 
but its statistical significance is hard to assess. For
example, J-type carbon stars clearly show \crat ratios smaller than 10,
well below the theoretical predictions ~\citep[e.g.][]{abi97,ohn99}.
Their progenitors appear to be
low-mass stars with $M\lesssim 2-3 M_\odot$, suggesting the operation of 
mixing processes ~\citep{abi00}. However, J-type represent only 
$\sim$15\% of the whole C-star population.  The occurrence of mixing in 
this class of stars finds support from the analysis of $s$-process elements 
in a sample of C(N)-type stars \citep{abi01}. 
These stars account for the majority of the C-stars
and have values of \crat greater than 10 but less than the solar value
(89), consistent with model results only for $M\lesssim 1.5 M_\odot$
(cf. Fig.~\ref{fig:abund}). 
For $M\gtrsim 1.5 M_\odot$, standard models predict
a sharp rise of the \crat ratio (up to $\sim$100), which is quite sensitive
to metallicity effects. The reduction of this ratio induced by the occurrence
of deep mixing in the AGB phase cannot be quantified at present, and 
detailed modelling is obviously needed to clarify the situation.

Owing to the lack of non standard AGB yields, we
show in the bottom right panel of Fig.~\ref{fig:abund} the results for RGB
deep mixing. How representative are these numbers for the more advanced 
phases? From our previous estimate of the \crat ratio in 20 PNe
from CO line emission, we have found that the majority of them have \crat
ratios between 10 and 25 for masses 1.5 to 3~$M_\odot$, These values are
lower than those expected from standard AGB models, even in the case where
the \crat value was very low ($\sim$10) at the base of the AGB, as a result 
of deep mixing
during the RGB. The strong \cdod production by third dredge-up along the AGB
raises the isotopic ratio to much higher values \citep{abi01}.  Therefore,
the overall evidence seems to indicate that active mixing occurs also in the
early AGB phase.  Returning to the case of NGC~3242, we may conclude that it
must belong to the small group of ``standard'' low-mass stars, consisting of
$\sim$10\% of stars with $M\lesssim 2$~\msun.

This point is of direct relevance to the \etre probem in the context of
Galactic chemical evolution.  A very small percentage of low-mass stars
producing large quantities of \etre is required to explain the \etre
abundances measured in the pre-solar nebula, in the local ISM, and in H{\sc
ii} regions ~\citep{pal00}.  All these objects show \heh ratios around
$10^{-5}$, lower than that measured in NGC~3242 by at least an order of
magnitude.

\begin{figure}
\plotone{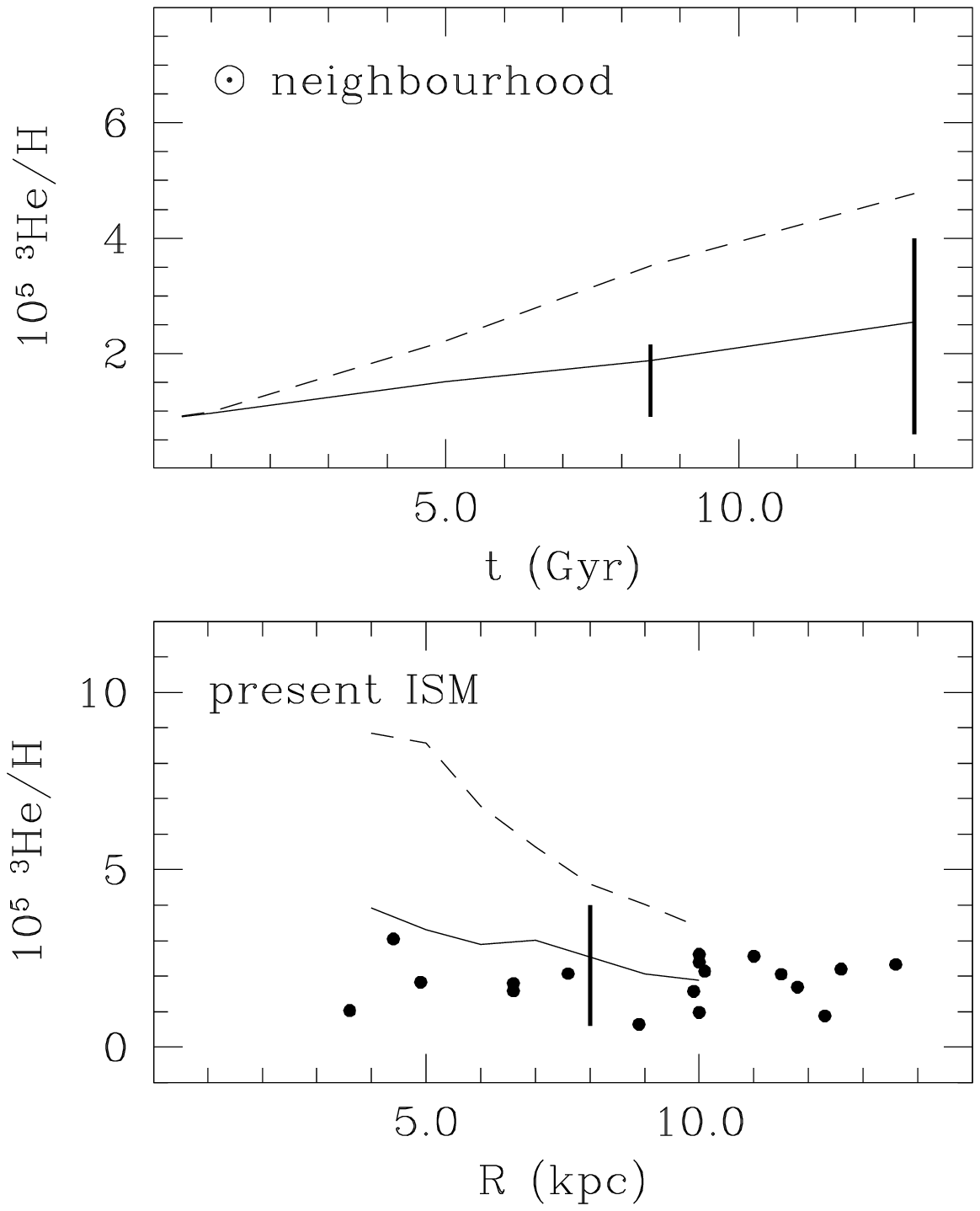}
\caption{Top panel: Galactic chemical evolution of \etre in the solar
neighborhood assuming (\etre/H)$_0=0.9\times 10^{-5}$
\citep{pal00,tos01}. The {\it dashed} curve is for a model
with standard yields for all stars, while the solid
curve corresponds to the same model but adopting the extra-mixing
yields by \citet{sac99} in 90\% of low-mass stars. The vertical bars
show the 2$\sigma$ \heh abundance for the local ISM and the solar
system \citep{gei98}.  Bottom panel: Radial distribution
of the \etre abundance at the present epoch. The data points display
the values of Galactic H{\sc ii} regions \citep{ban02}, while the
vertical bar at 8 kpc is the local ISM value as in the top panel. Line
symbols as above.
}
\label{fig:evol} 
\end{figure}

The upper panel of Figure~\ref{fig:evol} shows the predictions of chemical
evolution models assuming standard and non standard stellar nucleosynthesis
and the most recent abundances measured in the solar system and local ISM.
The model computed with the extra-mixing prescriptions of \citet{sac99} for
90\% of low-mass stars fits very well the observations, whereas the standard
model is systematically higher.

The failure of standard models is more dramatic when considering the
Galactic radial distribution of \etre, as shown in the bottom panel of 
Fig.~\ref{fig:evol}. The mixing hypothesis improves significantly the
agreement, and provides an upper envelope to the
distribution of {\it simple} H{\sc ii} regions, i.e. those which are
well described by homogeneous spheres 
\citep{ban02}. Bania et al. have argued that only these objects 
provide sufficiently accurate \heh estimates for
comparison with Galactic evolution models. 
Since {\it simple} H{\sc ii} regions have
\etre abundances systematically lower than the average, the implication is
that almost 100\% of low-mass stars should undergo non-standard 
nucleosynthesis. In that event, NGC~3242 would stand as a unique object in
the Galaxy.

\acknowledgments

The research of D.G., F.P. and M.T. has been supported by grants 
COFIN~1998-MURST and COFIN~2000-MURST at the Osservatorio Astrofisico di 
Arcetri and Osservatorio Astronomico di Bologna.

\clearpage

\clearpage

\end{document}